\documentclass[
reprint,
amsmath,
amssymb,
aps,
]{revtex4-2}

\usepackage{graphicx}
\usepackage{appendix}

\usepackage{subfig}
\usepackage{dcolumn}
\usepackage{bm}

\usepackage[mathlines]{lineno}

\usepackage{xcolor}
\definecolor{navyblue}{RGB}{0,0,150}
\usepackage[colorlinks, citecolor=purple,anchorcolor=purple,menucolor=purple, linkcolor=purple,filecolor=purple,runcolor=purple,urlcolor=purple,frenchlinks=purple, urlcolor=purple]{hyperref}

\usepackage{varioref}
\usepackage{cleveref}
\usepackage{orcidlink}
\usepackage{bbold}
\begin{document}
	
	\title{\color{navyblue}{SU($2$) string tension in the continuum limit using an effective theory of center vortices}}
		\author{Z.~Asmaee$^{1}$\orcidlink{0000-0002-3357-0574}}
	\email{zahra.asmaee@ut.ac.ir}
	\thanks{Corresponding author}

	\author{M.~Kiamari$^{2}$\orcidlink{0000-0001-9573-4020}}
\email{mkiamari@ipm.ir}

	\author{S.~Deldar$^{1}$\orcidlink{0000-0002-3959-514X}}
\email{sdeldar@ut.ac.ir}
\thanks{Corresponding author}

	\affiliation{
	$^{1}$Department of Physics, \href{https://ut.ac.ir/en}{University of Tehran}, North Karegar Avenue, Tehran 14395-547, Iran\\
	$^{2}$School of Particles and Accelerators, \href{https://particles.ipm.ir/}{Institute for Research in Fundamental Sciences (IPM)}, Tehran 19395-5531, Iran}	
	\begin{abstract}
		Using an effective theory for an ensemble of center vortices, we observe the area law fall-off in the continuum limit for the SU($2$) gauge group in three-dimensional Euclidean space-time. The string tension is obtained in terms of the intrinsic properties of the vortices and a parameter which describes their interactions. In addition, fitting our analytical results on lattice data, we show that the repulsive force between the vortices increases with temperature. Increasing the repulsive forces between the vortices at higher temperatures may prevent the vortices from having stable topological structures required for quark confinement.
		\begin{description}
			\item[PACS numbers] 74.25.Ha, 74.25.Uv, 12.38.Aw, 12.38.Lg.
		\end{description} 
	\end{abstract}
	\pacs{Valid PACS appear here}
	\maketitle
	
	\section{\label{sec:level1}Introduction}

	In the infrared regime of Quantum Chromodynamics (QCD), the strong coupling constant becomes large, leading to non-perturbative phenomena such as the color confinement mechanism, which remains one of the most controversial unsolved problems in particle physics  \cite{confinement2004, Greensite}.
	
	Lattice QCD and phenomenological models can be introduced as the most popular non-perturbative methods to explain the quark confinement \cite{Ichie, deldar, Deldar1}.
	The attempt to clarify the mechanism of the confinement phenomenon has led to the fact that the QCD vacuum has some non-trivial structure and this structure is responsible for the confinement of quarks inside the hadrons \cite{Ichie,bruckmann}.
	Topological objects like magnetic monopoles, vortices, instantons, dyons, and calorons are among the candidates that can be used to describe the confinement via some phenomenological models \cite{Ichie, Deldar2, Deldar3, Deldar4}. 
	
	From lattice QCD simulations, the area law fall-off for the Wilson loop average, in the infrared regime, is observed and it shows the linear potential that increases with the distance and is proportional to the string tension\cite{Ichie}.
	
	In the absence of matter fields, the center vortex model is a promising scenario for quark confinement.
	Historically, vortex-like structures were introduced in superconductors in 1959 \cite{Gorkov}, and recognized a few years later by Abrikosov \cite{Abrikosov}.
	It was proposed in various forms in the late 1970s with a field theoretical approach, for a good review, see \cite{kondo} and references therein. The idea is that the QCD vacuum is filled with closed magnetic vortices which can be condensed in the confined phase. 
	
	Lattice calculations show that vortices produce full string tension as the Yang-Mills vacuum does. This is an encouraging motivation to study confinement via center vortices. If the center vortices are removed from the lattice, the string tension also disappears \cite{DFG97,DFG98}.
	For SU($2$) gauge group, the string tension obtained from the projected thin center-vortex ensemble reproduces almost $97.7\%$ of the fundamental string tension calculated from the non gauge fixed situation.
	
	The most common methods of identifying vortices in the lattice simulation are direct maximal center gauge (DMCG) \cite{DFG98} and indirect maximal center gauge (IMCG)  \cite{DFG97}.
	Using phenomenological models, vortices have been identified by different methods, as we have also conducted some studies on identifying vortices in the continuum limit of QCD for the SU($2$) and SU($3$) gauge group	\cite{asmaee,karimimanesh}. 
	
	After a brief review on center vortices in Sec. \eqref{subsec:vortex}, we discuss the partition function of an ensemble of vortices in Sec. \eqref{subsec:partition function} which was obtained by a technique that has long been used by polymer physicists.
	The idea is that a condensate of oriented closed strings can be mapped onto a complex scalar field in a field theoretical approach \cite{samuel, stone, stone78, samuel78, oxman}.
	Assuming the intrinsic properties for the center vortex loops such as the stiffness and the tension, as well as defining an interaction between the vortices, an effective partition function for an ensemble of vortices was obtained.
	
	In Sec. \eqref{sec:level4}, Motivated by Lattice QCD and phenomenological models that suggest vortices as a possible candidate for confinement mechanism, we calculate the Wilson loop average using an ensemble of closed center vortices in three-dimensional Euclidean space-time. 
	Unlike the analytical lattice treatment based on the Villain formulation developed by Oxman and Reinhardt in Ref.~\cite{oxman}, the present work investigates the same problem in the continuum limit.
	In Ref.~\cite{oxman}, the Wilson-loop area law and the corresponding string tension were derived within a lattice formulation of the effective center vortex theory. Starting from a lattice representation of the effective partition function, the theory was mapped onto a frustrated three-dimensional XY model, whose effective vortex ensemble was analyzed using the Villain approximation. As a result, the derivation of the area law and the extraction of the string tension were carried out entirely within the lattice approach.
    In the present work, we address the same physical problem from the continuum perspective. Our analysis starts from the effective continuum partition function introduced in Ref.~\cite{oxman}. Instead of formulating the analysis on the lattice, we investigate the continuum realization of the effective theory and derive the Wilson-loop area law directly in the continuum limit. In doing so, we encounter a compact scalar field whose compactness cannot be treated within a purely continuum description. To properly account for this compact degree of freedom and its associated topological contributions, we employ a discrete representation only as an intermediate step. After performing the required transformations, the continuum formulation is fully recovered, and all physical results are derived within this framework.
	 We show that the Wilson loop average displays an area law fall-off and the string tension is derived for the SU($2$) gauge group in terms of the intrinsic properties of the center vortices and a parameter which describes the interaction between vortices. By fitting our analytical results, particularly the ratio of vortex stiffness to tension, to lattice data and using the temperature-dependent SU(2) string tension from Lattice QCD, we discuss the temperature dependence of the vortex interaction strength.

	\section{\label{sec:rev}Review Material }
	
	\subsection{\label{subsec:vortex}Center Vortices and Wilson loop}
	
	Center vortices are defined by the centers of the SU($N$) gauge group and they carry magnetic fluxes corresponding to the centers, $\text{Z($N$)}$. It is assumed that these magnetic fluxes squeeze the electric fields between a quark-antiquark pair inside a flux tube and as a result the confinement happens. If a Wilson loop is linked to a vortex in an SU($N$) gauge group, it gets a phase equal to the non-trivial center elements associated with $z$($k$), 
	\begin{equation}
		W(C)\longrightarrow z(k)W(C),
		\label{phase difference}
	\end{equation}
	where,
	\begin{equation}
		z(k)=e^{i\frac{2\pi}{N}k}\textbf{1}_{N \times N}, \quad\left( k=1, 2, ..., N-1\right).
		\label{center element}
	\end{equation}
  	As a result, center vortices emerge in the lattice, leading to an area-law behavior of the Wilson loop. 
	The center elements $z(k)$ can also be rewritten as the following \cite{oxman}
	\begin{equation}
		z(k)=e^{i2\pi \nu(k)}, \quad\nu(k)=\nu^a(k)H_a,
		\label{center element lie}
	\end{equation}
	where $\nu(k)$ are co-weights and $H_a$'s indicate the generators of Cartan sub-algebra.
	
\begin{figure}[ht]
		\begin{center}
			\centering
			\subfloat[]{\includegraphics[height=2.6cm, width=4.3cm]{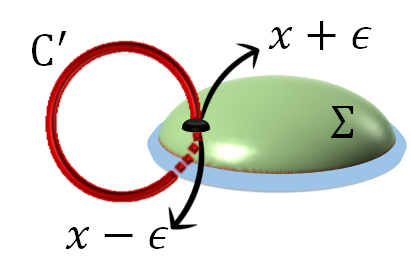}}
			\\
			\subfloat[]{\includegraphics[height=3cm, width=4cm]{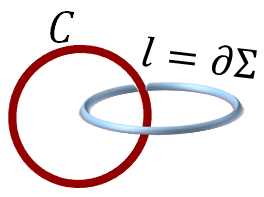}}
			\\
			\subfloat[]{\includegraphics[height=2.5cm, width=4.2cm]{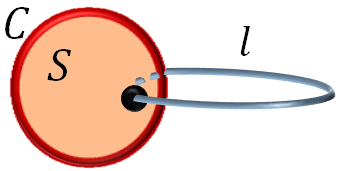}}
			\caption{(a) Intersection between the Wilson loop $C$ and hypersurface $\Sigma$. 
				(b) Linking between the Wilson loop $C$ and a vortex loop $l$ which is the boundary of hypersurface, $l=\partial\Sigma$. 
				(c) Intersection between the Wilson surface $S$ and the vortex loop $l$.}
			\label{fig1}
		\end{center}
	\end{figure}

	In three-dimensional space-time, a non-trivial contribution of a vortex to the Wilson loop operator is obtained by the linking number $L\left( C, l\right) $ between the Wilson loop $C$ and the vortex loop $l$, where $l$ indicates the boundary of the hypersurface $\Sigma$ $\left(\text{ i.e.}\ l=\partial \Sigma\right)$ shown in Fig. \ref{fig1}(b).
	The linking number can also be described in other alternative ways: by the intersection number $I\left( C,\Sigma\right) $ between the Wilson loop $C$ and the hypersurface $\Sigma$ shown in Fig. \ref{fig1}(a), or by the
	intersection number $I\left(S,l\right) $ between the Wilson surface $S$ and the vortex loop $l$ (see Fig. \ref{fig1}(c)). These three definitions are equivalent \cite{kondo}.
	
	We would like to calculate the Wilson loop average from the partition function of a vortex ensemble for SU($2$) gauge group to obtain the area law fall-off in an analytical way to represent quark confinement. 
	The contribution of the vortex field $A_\mu^l$ to the Wilson loop is as the following,
	\begin{equation}
		W_{\text{vortex}}(C)\equiv e^{i\oint_C dx_\mu A_\mu^l(x)}.
		\label{wilson loop of gauge field}
	\end{equation}
	On the other hand, when the thin vortex loop $l$ intersects the Wilson surface $S$, the Wilson loop obtains a factor as the following,
	\begin{equation}
		W_{\text{vortex}}(C)=z^{I(S,l)},
		\label{wilson loop of intersect}
	\end{equation}
	where $z$ is the non-trivial center element of the SU($2$) gauge group.
	Using Eqs. \eqref{center element lie}, \eqref{wilson loop of gauge field} and \eqref{wilson loop of intersect}, the relation between intersection number and the line integral of a vortex field $A_\mu^l$ is obtained,
	\begin{equation}
		\oint_C dx_\mu A_\mu^l(x)=2\pi\nu(k) I(S,l).
		\label{gauge field and I}
	\end{equation}
	For simplicity, we set $\nu(k)\equiv \nu$ in the subsequent calculations.
	This is justified since we work within an SU($2$)  gauge theory, where, according to Eq. \eqref{center element}, $k=1$, and thus no ambiguity arises from this simplification.
	An explicit integral formula for the intersection $I\left(S, l\right)$ between the Wilson surface $S$ and the vortex loop $l$ is defined \cite{kondo},
	\begin{equation}\begin{split}
			&I\left(S, l\right)=\dfrac{1}{4\pi}\oint_{l=\partial \Sigma}dx_{\mu}\int_{S}d^2{\sigma}_\mu\delta^{\left( 3\right) }\left(x-\bar{x}\left(\sigma \right)  \right),
			\\
			&d^2{\sigma}_\mu=\epsilon_{\mu\nu\rho}\dfrac{\partial \bar{x}_\nu}{\partial \sigma_1}\dfrac{\partial \bar{x}_\rho}{\partial \sigma_2}d\sigma_1d\sigma_2,
			\label{intersection formula}
	\end{split}\end{equation}
	where $d^2{\sigma}_\mu$ is the surface element and $S$ is parameterized by $\bar{x}\left(\sigma \right)= \bar{x}\left(\sigma_1,\sigma_2 \right)$.
	
	Replacing Eq. \eqref{intersection formula} in Eq. \eqref{gauge field and I}, and defining the vector field $J_\mu^C$ on the Wilson surface $S$
	\begin{equation}
		J_\mu^C(x)\equiv \dfrac{\nu}{2}\int_{S}d^2\sigma_\mu\delta^{\left( 3\right) }\left(x-\bar{x}\left(\sigma \right)  \right),
		\label{J}
	\end{equation}
	one obtains,
	\begin{equation}
		\oint_C dx_\mu A_\mu^l(x)=\oint_{l}dx_\mu J_\mu^C(x).
		\label{A and J}
	\end{equation}
	The linking number $L(C, l)$ is symmetric with respect to the interchange of the loops $C$ and $l$: $L(C, l)=L(l, C)$. Applying this interchange to the previous equation, one obtains a noticeable result. The vortex field $A_\mu^l$ and the vector field $J_\mu^l$ are somehow equivalent. The vortex field $A_\mu^l$ which is defined on the closed loop $l$ can be gauge transformed to the vector field $J_\mu^l$. Indeed, when a vortex intersects the Wilson surface, its flux passes through the surface $S$, therefore, the vector field $J_\mu^C$ indicates the gauge potential of the vortex field which its worldline $l$ is replaced by $C$ \cite{oxman}. Thus, the Wilson loop average is,
	\begin{equation}
		W_{\text{vortex}}(C)=e^{i\int_0^L ds u_\mu(s) J_\mu^C\left( x(s)\right)}.
		\label{W and J}
	\end{equation}
	where $L$ is the total length of the vortex worldline and $s$ is the arc length parameter of the vortex loop which runs from $0$ to $L$, and
	$u_\mu(s)$ is a tangent vector to the $s$ \cite{oxman}.

	For the ensemble of $n$ vortices intersecting the Wilson surface $S$,
	\begin{equation}
		W_{\text{vortex}}(C)=e^{i\sum_{i=1}^n\int_0^{L_i} ds_i u_\mu(s_i)  J_\mu^C(x(s_i))}.
		\label{W and J for n vortices}
	\end{equation}
	In fact, $J_\mu^C(x)$ describes a general vector field containing the flux of the $n$ vortices passing through surface $S$.
	
	\subsection{\label{subsec:partition function}Partition function for an ensemble of center vortices }
	
	To calculate the linear potential between a pair of quark-antiquark and the string tension in three-dimensional Euclidean space-time, we use the partition function of an ensemble of center vortices introduced in Ref. \cite{oxman}.
	
	The action of an ensemble of $n$ vortices is as the following \cite{oxman, stiff1, stiff2, stiff3},
	\begin{equation}
		S^0_n=\sum_{i=1}^n\int_0^{L_i}ds_i\left[\mu +\dfrac{1}{2\kappa}\left( \dfrac{du_\mu(s_i)}{ds_i}\right)^2\right], 
		\label{nonintraction action}
	\end{equation}
	where $\mu$ indicates the action per length and is called the tension of the center vortex. The parameter $\dfrac{1}{\kappa}$ shows the stiffness of the vortices and it is small for flexible vortices.
	The structure of a vortex is described by a space curve $x(s)$ in which $s\in \left[ 0, L\right] $ is a parameter denoting the arc length along the vortex backbone. 
	The vector $ u\left( s\right) = \frac{dx\left( s\right) }{ds}$ is a tangent vector and $\frac{du\left( s\right)}{ds}= \frac{d^2x\left( s\right) }{ds^2}$ is interpreted as the local curvature of the vortex. \cite{Fredrickson}
	
	To make the model more physical, one should assume a potential that represents the repulsive interaction between the vortices \cite{oxman},
		\begin{equation}\begin{split}
			&	S^{\text{int}}_n=\dfrac{1}{2}\sum_{i,j=1}^n\int_0^{L_i}\int_0^{L_j}ds_ids_jV(x_i(s_i),x_j(s_j)),
			\\
			&V(x_i-x_j)=\dfrac{1}{\zeta} \delta^{\left( 3\right)}(x_i-x_j),\quad \zeta>0,
			\label{interaction action}
	\end{split}\end{equation}
	where the parameter $\dfrac{1}{\zeta}$ may be interpreted as the strength of the potential.
	
	Therefore, to find the area law fall-off, one can define the Wilson loop average in terms of two partition functions,
	\begin{equation}
		\left\langle W(C)\right\rangle =\dfrac{Z[J^C_\mu]}{Z[0]},
		\label{W and Z}
	\end{equation}
where the relation $Z[J^C_\mu=0]\equiv Z[0]$ corresponds to the case where the Wilson loop is decoupled from the vortex ensemble.
 $Z[J^C_\mu]$ is the partition function of an ensemble of $n$ vortices in the presence of the Wilson loop \cite{oxman}, 
	\begin{equation}
		Z\left[ J_\mu^C\right]=\sum_n \int \left[ Dl\right] _n e^{i\sum_{i=1}^n\int_0^{L_i} ds_iu_\mu(s_i) J^C_\mu(x(s_i))} e^{-S^{\text{total}}_n},
		\label{initial partition}
	\end{equation}
	and the measure $\left[ Dl\right] _n$ integrates over all possible states of $n$ vortices and $S_{n}^{total}=S_{n}^{0}+S_{n}^{int}$.
	
	Using polymer techniques, Oxman and Reinhardt obtained an effective partition function of the form \cite{oxman},
	\begin{equation}
		Z\left[J^C_\mu \right]=\int \left[ D\bar{V}\right]  \left[ DV\right] \ e^{-\int d^3x\left(\frac{1}{3\kappa} \bar{D}_\mu \bar{V} D_\mu V+\mu\bar{V}V+\frac{1}{2\zeta}\left( \bar{V}V\right) ^2 \right) },
		\label{final partition}
	\end{equation}
	where $V$ is a complex scalar field and $D_\mu=\partial_\mu-iJ^C_\mu$ denotes the covariant derivative associated with the Wilson loop current $J^C_\mu$ \cite{oxman}. In this framework, the ensemble of oriented closed strings is mapped onto a complex scalar field description \cite{oxman, Fredrickson, oxman1}.
	
	In the next section, we use the effective partition function Eq. \eqref{final partition} to calculate the Wilson loop average from which the string tension is extracted. The dependence of the string tension on the vortex parameters and the temperature are discussed in detail.
	
	\section{\label{sec:level4}the Wilson loop average in the presence of the center vortices - the area law fall-off }
	
	From the partition function in Eq.~\eqref{final partition}, the effective Lagrangian of the vortex ensemble is identified as
	\begin{equation}
		\mathcal{L}=\dfrac{1}{3\kappa} \bar{D}_\mu \bar{V} D_\mu V+\mu\bar{V}V+\dfrac{1}{2\zeta}\left( \bar{V}V\right) ^2.
		\label{lagrangian}
	\end{equation}
The Lagrangian $\mathcal{L}$ has a global $U(1)$ symmetry. The potential term can be rewritten in the vacuum form, leading to
		\begin{equation}
		Z\left[J^C_\mu \right]=\int \left[ D\bar{V}\right]  \left[ DV\right] \ e^{-\int d^3x\left(\frac{1}{3\kappa} \bar{D}_\mu \bar{V} D_\mu V+\frac{1}{2\zeta}\left(\bar{V}V-v^2 \right) ^2 \right) },
		\label{partition, vacuum}
	\end{equation}
	where $	v^2=-\mu\zeta>0, \quad \mu<0$.
	To condense the center vortices, one should have $\bar{V}V=v^{2}$. Hence the complex scalar field $V(x)$ can be defined as,
	\begin{equation}
		V(x)=\rho(x)e^{i\gamma(x)}, \quad \rho(x)=v+h(x),
		\label{complex field}
	\end{equation}
	where $h(x)$ represents the radial fluctuation around the vacuum expectation value $v$.
	For sufficiently weak vortex interactions, the potential in Eq.~\eqref{partition, vacuum} induces only small fluctuations in the field $h(x)$, which can be neglected compared to the vacuum expectation value $v$ \cite{oxman}. 
	This approximation is valid in the weak-interaction regime characterized by $|\mu|\zeta\gg1$, where the interaction strength controlled by $1/\zeta$ is small compared with the characteristic energy scale $|\mu|$ associated with the center vortex tension. 
	In this regime, one has $v^2=|\mu|\zeta\gg1$, and therefore the radial fluctuations are suppressed relative to the vacuum expectation value, making the mean-field approximation $\rho(x)\simeq v$ consistent. 
	Moreover, expanding the potential around its minimum gives a mass for the radial mode,
	$
	m_h^2=4|\mu|,
	$
	which shows that the radial degree of freedom becomes increasingly suppressed for large $|\mu|$. 
	Therefore, its contribution to the string tension appears only through subleading corrections beyond the approximation considered here.
	The compact scalar field $\gamma(x)$ describes the remaining phase fluctuations and is a modulo $2\pi$ function.
	
	Using Eq.~\eqref{complex field} and replacing the covariant derivative with the partial derivative, the partition function $Z[0]$ takes the form
	\begin{equation}
	Z[0]=N^\prime v^2\int_{-\pi} ^{\pi}\left[ D\gamma(x)\right]  e^{-\frac{v^2}{3\kappa}\int d^3x\left(\partial_\mu\gamma(x)\right)  ^2},
	\label{Z02}
	\end{equation}
	where $N^\prime$ denotes a normalization factor. In the next two subsections, we present a detailed analytical calculation of the partition function of Eq. \eqref{Z02}, followed by its generalization in the presence of Wilson loop.
	
	\subsection{\label{subsec:compact field}Integrating on compact scalar field}
	
	The partition function of Eq. \eqref{Z02} can be calculated by the Fourier expansion for a non-compact scalar field. But $\gamma(x)$ is a compact scalar field which lives on a circle.
	 Furthermore, given the definition of the complex field in Eq. \eqref{complex field}, the partition function of Eq. \eqref{Z02} is not exactly equivalent to Eq. \eqref{partition, vacuum} when $J_{\mu}^{C} = 0$: while the latter remains invariant under the transformation $\gamma(x) \rightarrow \gamma(x) + 2\pi n$ ($n \in \mathbb{Z}$), the former does not.
	 The compact nature of the scalar field $\gamma(x)$ prevents a direct treatment within a purely continuum formulation. Therefore, following the procedure introduced in Ref.~\cite{Polyakov-book}, we temporarily employ a discrete representation to properly account for the compact scalar sector. This discrete formulation serves only as an intermediate step.

	An effective approach to solving the periodic Gaussian integral \eqref{Z02} is to employ a discrete formulation in which $\gamma_{\boldsymbol{x}}$ is allowed to be a multivalued function. This procedure, commonly implemented through the Villain formulation, provides a convenient way to account for the compact nature of the scalar field \cite{Polyakov-book}.
	This goal can be achieved by substituting Eq. \eqref{Z02} with:
	\begin{equation}
	Z[0]=N^\prime v^2\sum_{\left\lbrace n_{\boldsymbol{x},\boldsymbol{\delta}}\right\rbrace }\int_{-\pi}^{\pi}  [D\gamma_{\boldsymbol{x}}]e^{-\frac{v^2}{3\kappa}a\sum_{\boldsymbol{x},\boldsymbol{\delta}}\left(  \gamma_{\boldsymbol{x}}-\gamma_{\boldsymbol{x}+\boldsymbol{\delta}}+2\pi n_{\boldsymbol{x},\boldsymbol{\delta}}\right)^2},
	\label{Z0 lattice}
	\end{equation}
	where The parameter $a$ introduced in the discrete representation plays the role of a discretization scale.
	$n_{\boldsymbol{x},\boldsymbol{\delta}}$ is an integer-valued plaquette variable introduced to account for the compactness of the scalar field. These variables provide a convenient representation of the different topological sectors of the compact scalar field. The integer-valued variables $n_{\boldsymbol{x},\boldsymbol{\delta}}$ contain two contributions: one associated with the gauge field residing on the links and the other associated with the string configuration penetrating the plaquette \cite{Polyakov-book, shnir}.
	 They can be decomposed as
	\begin{equation}
	n_{\boldsymbol{x},\boldsymbol{\delta}}=m_{\boldsymbol{x}}-m_{\boldsymbol{x}+\boldsymbol{\delta}}+\alpha_{\boldsymbol{x}}-\alpha_{\boldsymbol{x}+\boldsymbol{\delta}}+\varepsilon_{\boldsymbol{\delta}\lambda}\left(\phi_{\boldsymbol{x}_*}-\phi_{\boldsymbol{x}_*-\lambda} \right),
	\label{n}
	\end{equation}
	where $m_{\boldsymbol{x}}$ and $\left| \alpha_{\boldsymbol{x}}\right| <1$ are integer valued vector fields defined on the links of the square lattice. $\boldsymbol{x}_*$ denotes the centers of the plaquettes, and $\varepsilon_{\boldsymbol{\delta}\lambda}$ is the standard antisymmetric tensor. The variables $\phi_{\boldsymbol{x}_*}$ represent a set of integer-valued variables defined at the centers of the plaquettes. The discrete Laplacian of this representation leads to an integer-valued quantity that characterizes the topological structure of the compact scalar field. This quantity satisfies the following relation \cite{Polyakov-book}:
	\begin{equation}
	\bigtriangleup_{\boldsymbol{x}_*, \boldsymbol{x}_*^\prime}\phi_{\boldsymbol{x}_*^\prime}\equiv  \sum_{\lambda}\left(\phi_{\boldsymbol{x}_*-\lambda} +\phi_{\boldsymbol{x}_*+\lambda}-2\phi_{\boldsymbol{x}_*}\right)=q_{\boldsymbol{x}_*},
	\label{laplacian}
	\end{equation}
	here, the integers $q_{\boldsymbol{x}_*}$ represent the field strength generated by the vector potential $n_{\boldsymbol{x},\boldsymbol{\delta}}$.
Using Eqs.~\eqref{n} and \eqref{laplacian}, the summation over the discrete variables $\left\lbrace n_{\boldsymbol{x},\boldsymbol{\delta}} \right\rbrace$ is transformed into the summations over the integer-valued variables $\left\lbrace m_{\boldsymbol{x}} \right\rbrace$ and $\left\lbrace q_{\boldsymbol{x}_*} \right\rbrace$. 

By substituting Eq. \eqref{n} into Eq. \eqref{Z0 lattice} and performing a change of variables as,
	\begin{equation}
	\gamma_{\boldsymbol{x}}\rightarrow \gamma_{\boldsymbol{x}}-2\pi\left( m_{\boldsymbol{x}}+\alpha_{\boldsymbol{x}}\right), 
	\label{change}
	\end{equation}
	Eq.~\eqref{Z0 lattice} can be rewritten as \cite{Polyakov-book},
	\begin{equation}\begin{split}
			Z[0]
			&=\sum_{\left\lbrace q_{\boldsymbol{x}_*},q_{\boldsymbol{x}^\prime_*}\right\rbrace }e^{- i\frac{2\pi^2v^2}{3\kappa} a^3\sum_{{\boldsymbol{x}_*},\boldsymbol{x}^\prime_*} q_{\boldsymbol{x}_*}\bigtriangleup_{\boldsymbol{x}_*,\boldsymbol{x}_*^\prime}^{-1}q_{\boldsymbol{x}_*^\prime}}\\
			&	\times \int_{-\infty}^{\infty}  [D\gamma_{\boldsymbol{x}}]e^{-\frac{v^2}{3\kappa}a \sum_{\boldsymbol{x},\boldsymbol{\delta}}\left(  \gamma_{\boldsymbol{x}}-\gamma_{\boldsymbol{x}+\boldsymbol{\delta}}\right) ^2}.
			\label{Z0 lattice3}
				\end{split}
	\end{equation}
	The integral in Eq.~\eqref{Z0 lattice3} is Gaussian. After the above change of variables, the compact degrees of freedom are transferred to the integer-valued variables, and $\gamma_{\boldsymbol{x}}$ can be treated as a non-compact scalar field. Therefore, the integral can be evaluated using the Fourier expansion, as will be shown in the next subsection.

	\subsection{\label{subsec:ZJ}The area law fall-off}

In this subsection, we calculate the partition function $Z[J^C_\mu]$ in Eq.~\eqref{partition, vacuum} when the Wilson loop links the vortices. Using the definition of the covariant derivative, the path integral weight can then be represented as a product of four exponential factors,
	\begin{equation}
		\begin{split}
		Z[J^C_\mu]=&\int \left[ D\bar{V}\right] \left[DV\right] \ e^{-\frac{1}{3\kappa}\int d^3x \partial_\mu\bar{V}(x)\partial_\mu V(x)}\\
		&\times e^{-\frac{1}{3\kappa}\int d^3x J^C_\mu \bar{V}(x)J^C_\mu V(x)}\, e^{\frac{i}{3\kappa}\int d^3x \left( \partial_\mu \bar{V}(x)\right)J^C_\mu V(x) }\\
		&\times e^{-\frac{i}{3\kappa}\int d^3xJ^C_\mu(x)\bar{V}(x)\partial_\mu V(x)}.
		\end{split}
		\label{ZJ1}
	\end{equation}
Each exponential factor in Eq.~\eqref{ZJ1} will be evaluated separately in the following.

\vspace{0.3cm}
\textbf{(a):} The first exponential factor in Eq.~\eqref{ZJ1} is given by,
	\begin{equation}\begin{split}
		I_1&\equiv e^{-\frac{1}{3\kappa}\int d^3x \partial_\mu\bar{V}(x)\partial_\mu V(x)}
		\\
		&=e^{-\frac{v^2}{3\kappa}a \sum_{\boldsymbol{x},\boldsymbol{\delta}}\left(  \gamma_{\boldsymbol{x}}-\gamma_{\boldsymbol{x}+\boldsymbol{\delta}}\right) ^2 }
		e^{- i\frac{2\pi^2v^2}{3\kappa} a^3\sum_{\boldsymbol{x}_*,\boldsymbol{x}^\prime_*} q_{\boldsymbol{x}_*}\bigtriangleup^{-1}_{\boldsymbol{x}_*,\boldsymbol{x}_*^\prime}q_{\boldsymbol{x}_*^\prime}}.
		\label{first exponential}
	\end{split}\end{equation}
	In the second equality, we have used the results of the computations in Eq. \eqref{Z0 lattice3}.

\begin{figure}[ht]
	\begin{center}
		\centering
		\includegraphics[height=6.5cm, width=7.3cm]{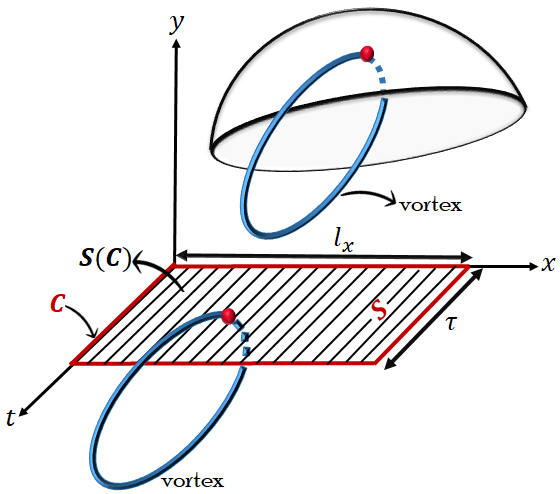}
		\caption{Intersection between the Wilson surface and the vortex loop.}
		\label{fig4}
	\end{center}
\end{figure}

\vspace{0.3cm}
	\textbf{(b):} The second exponential factor appearing in Eq.~\eqref{ZJ1} reads,
	\begin{equation}
	I_2\equiv e^{-\frac{1}{3\kappa}\int d^3x J^C_\mu \bar{V}(x)J^C_\mu V(x)}.
	\label{second exponential}
	\end{equation}
	The Wilson surface whose boundary is the Wilson loop, is not unique and different Wilson surfaces are related to each other by gauge transformations. Since, the string tension calculated from the Wilson loop average is a gauge invariant quantity and independent of the choice of the Wilson surface shape, we choose the simplest form. We consider the Wilson surface as a rectangular shape on the $t-x$ plane and each point on this Wilson surface is parameterized by $\left( \bar{t}, \bar{x}\right)$.
	 Fig. \ref{fig4} shows a schematic picture of a three dimensional Wilson surface which is projected on $t-x$ plane. The origin is located on one of the corners of the rectangle. 
	 
	Substituting Eqs.~\eqref{J} and \eqref{complex field} into Eq.~\eqref{second exponential}, we obtain
\begin{equation}\begin{split}
		I_2&=e^{-\frac{v^2 \nu^2}{12\kappa}\int_{S}d^2\sigma_\mu\int_{S}d^2\sigma^{\prime}_\mu\int d^3x \delta^{\left( 3\right)}\left(x- \bar{x}(\sigma)\right) \delta^{\left( 3\right)}\left(x-\bar{x}^{\prime}(\sigma^\prime)\right)}\\
		&=e^{-\frac{v^2 \nu^2}{12\kappa}\int_{S}d^2\sigma_\mu\int_{S}d^2\sigma^{\prime}_\mu \delta^{\left( 3\right)}\left( \bar{x}^{\prime}(\sigma^\prime)-\bar{x}(\sigma)\right)}\\
		&=e^{-\frac{v^2 \nu^2}{12\kappa}\int_{S}d^2\sigma_\mu\int_{S}d\bar{t}^{\prime}d\bar{x}^{\prime} \delta\left( \bar{t}^{\prime}-\bar{t}\right)\delta\left( \bar{x}^{\prime}-\bar{x}\right)\delta\left( \bar{y}^{\prime}-\bar{y}\right)}\\
		&=e^{-\frac{v^2 \nu^2}{12\kappa}\int_{S}d\bar{t} d\bar{x} \delta\left( \bar{y}^{\prime}-\bar{y}\right)}.
		\label{second exponential1}
	\end{split}\end{equation}
	We have chosen $\bar{y}^\prime=0$ since the projected Wilson loop is indeed located at the $t-x$ plane. Therefore, Eq.~\eqref{second exponential1} takes the form
	\begin{equation}
	I_2=e^{-\frac{ v^2 \nu^2}{12\kappa}\delta\left( \bar{y}\right) \int_0^\tau d\bar{t}\int_0^{l_x}d\bar{x}}
	=e^{-\frac{v^2 \nu^2}{12\kappa}\delta\left( \bar{y}\right)\times S},
	\label{final second exponential}
	\end{equation}
	where $S\equiv \tau l_x$ is the area of the Wilson loop in the $t-x$ plane. The term $\delta\left(\bar{y}\right)$ specifies the intersection location of the Wilson surface and the vortex loop (see Fig.~\ref{fig1}.(c)), indicating a localized contribution at $\bar{y}=0$.

\vspace{0.3cm}
\textbf{(c):} The third exponential factor in Eq.~\eqref{ZJ1} takes the form,
	\begin{equation}
	I_3\equiv e^{\frac{i}{3\kappa}\int d^3x \left( \partial_\mu \bar{V}(x)\right)J^C_\mu V(x)}.
	\label{third exponential}
	\end{equation}
	Replacing Eqs. \eqref{J} and \eqref{complex field} in Eq. \eqref{third exponential}, 
	\begin{equation}\begin{split}
		I_3&=e^{\frac{v^2\nu}{6\kappa}\int_{S}d^2\sigma_{\mu}\int d^3x\partial_\mu\gamma(x)\delta^{\left( 3\right)}\left( x-\bar{x}(\sigma)\right)}\\
		&=e^{\frac{v^2\nu}{6\kappa}\int_{S}d^2\sigma_{\mu}\partial_\mu\gamma(\bar{x}(\sigma))}.
		\label{third exponential1}
	\end{split}\end{equation}
	To calculate Eq.~\eqref{third exponential1}, we first rewrite it in a discrete representation, where $\gamma(\bar{x})$ is treated as a multivalued function. Accordingly, Eq.~\eqref{third exponential1} takes the form
	\begin{equation}\begin{split}
		I_3=e^{\frac{v^2\nu a}{6\kappa}\sum_{\bar{\boldsymbol{x}},\boldsymbol{\delta}}\left( \gamma_{\bar{\boldsymbol{x}}}-\gamma_{\bar{\boldsymbol{x}}+\boldsymbol{\delta}}+2\pi n_{\bar{\boldsymbol{x}},\boldsymbol{\delta}}\right)}.
		\label{third exponential2}
	\end{split}
	\end{equation}
	Substituting Eq.~\eqref{n} into Eq.~\eqref{third exponential2} and using Eq.~\eqref{change}, we obtain
	\begin{equation}\begin{split}
		I_3=e^{\frac{v^2\nu}{6\kappa}a\sum_{\bar{\boldsymbol{x}},\boldsymbol{\delta}} \left( \gamma_{\bar{\boldsymbol{x}}}-\gamma_{\bar{\boldsymbol{x}}+\boldsymbol{\delta}}\right)} e^{\frac{2\pi v^2\nu}{6\kappa}a\sum_{\bar{\boldsymbol{x}}_*,\boldsymbol{\delta},\lambda}
			\epsilon_{\boldsymbol{\delta}\lambda}\left(\phi_{\bar{\boldsymbol{x}}_*}-\phi_{\bar{\boldsymbol{x}}_*-\lambda} \right) }.
		\label{third exponential3}
	\end{split}
	\end{equation}

	\vspace{0.3cm}
	\textbf{(d):} The fourth exponential factor in Eq.~\eqref{ZJ1} can be written as,
	\begin{equation}
	I_4\equiv e^{-\frac{i}{3\kappa}\int d^3xJ^C_\mu(x)\bar{V}(x)\partial_\mu V(x)}.
	\label{fourth exponential}
	\end{equation}
	Substituting Eqs.~\eqref{J} and \eqref{complex field} into Eq.~\eqref{fourth exponential}, and following the same procedure used in part \textbf{(c)}, this contribution is rewritten in a discrete representation as
	\begin{equation}\begin{split}
			I_4&=e^{\frac{v^2\nu}{6\kappa}\int_{S}d^2\sigma_{\mu}\int d^3x\partial_\mu\gamma(x)\delta^{\left( 3\right)}\left( x-\bar{x}(\sigma)\right)}\\
			&=e^{\frac{v^2\nu}{6\kappa}\int_{S}d^2\sigma_{\mu}\partial_\mu\gamma(\bar{x}(\sigma))}.
			\label{fourth exponential1}
	\end{split}\end{equation}
	Since Eq.~\eqref{fourth exponential1} is identical to Eq.~\eqref{third exponential1}, the subsequent derivation follows exactly the same steps as those leading from Eqs.~\eqref{third exponential1} to \eqref{third exponential3}. Therefore,
	\begin{equation}
		I_4=
		e^{\frac{v^2\nu}{6\kappa}a\sum_{\bar{\boldsymbol{x}},\boldsymbol{\delta}}
			(\gamma_{\bar{\boldsymbol{x}}}-\gamma_{\bar{\boldsymbol{x}}+\boldsymbol{\delta}})}
		e^{\frac{2\pi v^2\nu}{6\kappa}a
			\sum_{\bar{\boldsymbol{x}}_*,\boldsymbol{\delta},\lambda}
			\varepsilon_{\boldsymbol{\delta}\lambda}
			(\phi_{\bar{\boldsymbol{x}}_*}-\phi_{\bar{\boldsymbol{x}}_*-\lambda}) }.
		\label{fourth exponential3}
	\end{equation}
	Now, we are ready to calculate the partition function for the Wilson loop linked with vortices using the results obtained above. Substituting Eqs.~\eqref{first exponential}, \eqref{final second exponential}, \eqref{third exponential3}, and \eqref{fourth exponential3} into Eq.~\eqref{ZJ1}, we obtain
	\begin{widetext}\begin{equation}\begin{split}
		Z[J^C_\mu]&=e^{-\frac{v^2 \nu^2}{12\kappa}\delta\left( \bar{y}\right)\times S}\sum_{\left\lbrace q_{\boldsymbol{x}_*},q_{\boldsymbol{x}_*^\prime}\right\rbrace }e^{- i\frac{2\pi^2v^2}{3\kappa} a^3\sum_{\boldsymbol{x}_*,\boldsymbol{x}^\prime_*} q_{\boldsymbol{x}_*}\bigtriangleup_{\boldsymbol{x}_*,\boldsymbol{x}_*^\prime}q_{\boldsymbol{x}_*^\prime}}
		 \sum_{\left\lbrace \phi_{\bar{\boldsymbol{x}}_*}\right\rbrace }e^{\frac{2\pi v^2\nu}{3\kappa}a\sum_{\bar{\boldsymbol{x}}_*,\boldsymbol{\delta},\lambda}
		 	\epsilon_{\boldsymbol{\delta}\lambda}\left(\phi_{\bar{\boldsymbol{x}}_*}-\phi_{\bar{\boldsymbol{x}}_*-\lambda} \right) }\\&
		\times N^\prime v^2\int_{-\infty}^\infty \left[D\gamma_{\boldsymbol{x}} \right] e^{-\frac{v^2}{3\kappa}a \sum_{\boldsymbol{x},\boldsymbol{\delta}}\left(  \gamma_{\boldsymbol{x}}-\gamma_{\boldsymbol{x}+\boldsymbol{\delta}}\right) ^2} 
		e^{\frac{v^2\nu a}{3\kappa}\sum_{\bar{\boldsymbol{x}},\boldsymbol{\delta}}\left( \gamma_{\bar{\boldsymbol{x}}}-\gamma_{\bar{\boldsymbol{x}}+\boldsymbol{\delta}}\right) },
		\label{ZJ2}
	\end{split}
	\end{equation}\end{widetext}
Next, substituting Eqs.~\eqref{Z0 lattice3} and \eqref{ZJ2} into Eq.~\eqref{W and Z}, the Wilson loop average for the vortex ensemble is obtained. After canceling the common factors in the numerator and denominator, we obtain
\newpage
	\begin{widetext}\begin{equation}
		\left\langle W(C)\right\rangle = e^{-\frac{v^2 \nu^2}{12\kappa}\delta\left( \bar{y}\right)\times S}\times N^\prime\sum_{\left\lbrace \phi_{\bar{\boldsymbol{x}}_*}\right\rbrace }e^{\frac{2\pi v^2\nu}{3\kappa}a\sum_{\bar{\boldsymbol{x}}_*,\boldsymbol{\delta},\lambda}
			\epsilon_{\boldsymbol{\delta}\lambda}\left(\phi_{\bar{\boldsymbol{x}}_*}
			-\phi_{\bar{\boldsymbol{x}}_*-\lambda} \right) }\times \dfrac{\int_{-\infty}^\infty \left[D\gamma_{\boldsymbol{x}} \right] e^{-\frac{v^2}{3\kappa}a \sum_{\boldsymbol{x},\boldsymbol{\delta}}\left(  \gamma_{\boldsymbol{x}}-\gamma_{\boldsymbol{x}+\boldsymbol{\delta}}\right) ^2} 
			e^{\frac{v^2\nu a}{3\kappa}\sum_{\bar{\boldsymbol{x}},\boldsymbol{\delta}}\left( \gamma_{\bar{\boldsymbol{x}}}-\gamma_{\bar{\boldsymbol{x}}+\boldsymbol{\delta}}\right) }}{N^\prime\, \int_{-\infty}^{\infty}  [D\gamma_{\boldsymbol{x}}]e^{-\frac{v^2}{3\kappa}a \sum_{\boldsymbol{x},\boldsymbol{\delta}}\left(  \gamma_{\boldsymbol{x}}-\gamma_{\boldsymbol{x}+\boldsymbol{\delta}}\right) ^2}}.
		\label{W2}
	\end{equation}\end{widetext}
At this stage, we proceed with the evaluation of Eq.~\eqref{W2}. To this end, we investigate the relation between $\phi_{\bar{\boldsymbol{x}}_*}$ and $q_{\bar{\boldsymbol{x}}_*}$ on the $t-x$ plane, where the Wilson surface is defined. In the two-dimensional space-time, the integer-valued variables $n_{\bar{\boldsymbol{x}},\boldsymbol{\delta}}$ can be represented in terms of the integers $q_{\bar{\boldsymbol{x}}_*}$ as follows \cite{Polyakov-book},
	\begin{equation}
	q_{\bar{\boldsymbol{x}}_*}=n_{\bar{\boldsymbol{x}},\mathbf{1}}
	+n_{\bar{\boldsymbol{x}}+\mathbf{1},\mathbf{2}}-n_{\bar{\boldsymbol{x}}+\mathbf{2},
		\mathbf{1}}
	-n_{\bar{\boldsymbol{x}},\mathbf{2}}=\sum_{\Box}n_{\bar{\boldsymbol{x}},\boldsymbol{\delta}}.
	\label{q1}
	\end{equation}
	In two-dimensional space-time, it is obtained by summing these variables around the boundary of the plaquette centered at $\bar{\boldsymbol{x}}_*$ with a chosen orientation. The alternating signs arise from the orientation of the plaquette boundary and ensure that only the net circulation of $n_{\bar{\boldsymbol{x}},\boldsymbol{\delta}}$ contributes to $q_{\bar{\boldsymbol{x}}_*}$. Consequently, $q_{\bar{\boldsymbol{x}}_*}$ measures the winding number enclosed by the plaquette and characterizes the topological sector of the compact scalar field.
	Substituting Eq.~\eqref{n} into Eq.~\eqref{q1}, the contributions containing $m_{\boldsymbol{x}}$ and $\alpha_{\boldsymbol{x}}$ cancel pairwise around the plaquette, while the remaining contribution is determined by the plaquette variable $\phi_{\boldsymbol{x}_*}$. This leads to
	\begin{equation}
	q_{\bar{\boldsymbol{x}}_*}=\sum_{\bar{\boldsymbol{x}}_*,\boldsymbol{\delta},\lambda}\epsilon_{\boldsymbol{\delta}\lambda}\left(\phi_{\bar{\boldsymbol{x}}_*}-\phi_{\bar{\boldsymbol{x}}_*-\lambda} \right).
	\label{q2}
	\end{equation}
Consequently, using Eq.~\eqref{q2}, the summation over ${\left\lbrace \phi_{\bar{\boldsymbol{x}}_*}\right\rbrace }$ in Eq.~\eqref{W2} is transformed into the summation over ${\left\lbrace q_{\bar{\boldsymbol{x}}_*}\right\rbrace }$. 
In addition, the second exponential factor in Eq.~\eqref{W2} can be rewritten in terms of $q_{\bar{\boldsymbol{x}}_*}$ as
\begin{align}
	e^{\frac{2\pi v^2\nu}{3\kappa}a\sum_{\bar{\boldsymbol{x}}_*,\boldsymbol{\delta},\lambda}
		\varepsilon_{\boldsymbol{\delta}\lambda}
		\left(\phi_{\bar{\boldsymbol{x}}_*}
		-\phi_{\bar{\boldsymbol{x}}_*-\lambda}\right)}
	=
	e^{\frac{2\pi v^2\nu}{3\kappa}a\sum_{\bar{\boldsymbol{x}}_*}
		q_{\bar{\boldsymbol{x}}_*}}.
\end{align}
Since $q_{\bar{\boldsymbol{x}}_*}$ is an integer-valued quantity, this exponential factor is independent of the continuous field $\gamma(x)$ and the Wilson surface geometry. Therefore, it contributes only to the overall normalization of the Wilson loop average. Since the factor independent of the Wilson surface area appears in the numerator of Eq.~\eqref{W2}, its contribution is absorbed into the normalization factor $N^\prime$ in the numerator, leading to a modified normalization constant. The ratio between this modified factor and the corresponding $N^\prime$ in the denominator of Eq.~\eqref{W2} gives an overall normalization factor, which can be omitted since it does not affect the area-law term or the extracted string tension.

After completing the treatment of the compact scalar field, we now return to the continuum formulation. Taking the continuum limit, $a\rightarrow0$, the discrete sums are replaced by the corresponding continuum integrals, while the discrete fields $\gamma_{\boldsymbol{x}}$ and $\gamma_{\bar{\boldsymbol{x}}}$ become the continuum fields $\gamma(x)$ and $\gamma(\bar{x})$, respectively, and the Wilson loop average takes the form:
	\begin{equation}\begin{split}
		\left\langle W(C)\right\rangle &=e^{-\frac{v^2 \nu^2}{12\kappa}\delta\left( \bar{y}\right)\times S}\\&\times \dfrac{\int_{-\infty}^\infty \left[D\gamma(x)\right] e^{-\frac{v^2}{3\kappa}\int d^3x\left( \partial_\mu\gamma(x)\right) ^2} 
			e^{\frac{v^2\nu }{3\kappa}\int d^2\sigma_\mu\partial_\mu\gamma({\bar{x}}) }}{\int_{-\infty}^{\infty}  [D\gamma(x)]e^{-\frac{v^2}{3\kappa}\int d^3x\left( \partial_\mu\gamma(x)\right) ^2}}.
		\label{W3}
	\end{split}\end{equation}
To calculate the ratio of the functional integrals in Eq.~\eqref{W3}, we rewrite the free part of the exponent as a quadratic action. In three-dimensional Euclidean space-time \cite{kapusta}, this action is given by
	\begin{equation}\begin{split}
		S_0&\equiv-\frac{v^2}{3\kappa}\int d^3x\left(\partial_\mu\gamma(x)\right)^2\\
		&=-\frac{v^2}{3\kappa}\int^\beta_0 dt\int  d^2x \gamma(x)\left(-\frac{\partial^2}{\partial t^2}-\nabla^2 \right) \gamma(x).
		\label{Sprime}
	\end{split}\end{equation}
Using the Fourier expansion of the scalar field $\gamma(x)$ \cite{kapusta},
	\begin{equation}
	\gamma(x)=\gamma\left(t,x,y \right)=\gamma\left( t, \vec{x}\right)=\sum_{n=-\infty}^\infty \sum_{\vec{p}} e^{i\left( w_nt+\vec{p}.\vec{x}\right) }\gamma_n(\vec{p}),
	\label{Fourier }
	\end{equation}
	where $w_n=2\pi nT$ denotes the Matsubara frequency and $T$ is the temperature. Substituting Eq.~\eqref{Fourier } into Eq.~\eqref{Sprime}, the quadratic action becomes	
	\begin{equation}
	S_0=-\dfrac{v^2\mathtt{V}}{3\kappa} \sum_{n=-\infty}^{\infty} \sum_{\vec{p}}\left( w_n^2+\vec{p}^2\right)\left| \gamma_n(\vec{p})\right| ^2 ,
	\label{Sprime1}
	\end{equation}
	where $\vec{p}^{\,2}=p_x^2+p_y^2$ and $\mathtt{V}\equiv\beta L^2$ is the volume of the three-dimensional Euclidean space-time. Here, $L$ and $\beta$ represent the spatial length and the extent of the imaginary time direction, respectively.
	Using the above representation, the Gaussian functional integral in Eq.~\eqref{W3} can be evaluated as
	\begin{equation}
	e^{S_0}=\prod_{n=-\infty}^{\infty} \prod_{\vec{p}} e^{-\frac{v^2\mathtt{V}}{3\kappa}\left( w_n^2+\vec{p}^2\right)\left| \gamma_n(\vec{p})\right| ^2}.
	\label{e1}
\end{equation}
The remaining exponential factor in Eq.~\eqref{W3} involves the coupling between the scalar field and the Wilson surface. For the chosen orientation of the Wilson surface, we have
	\begin{equation}
	d^2\sigma_\mu \partial_\mu\gamma(x(\sigma))=d^2\sigma_2 \partial_2 \gamma(x(\sigma))=d\bar{t}d\bar{x}\dfrac{\partial \gamma(x)}{\partial y},
	\label{sigma and gamma}
	\end{equation}
where, according to Fig.~\eqref{fig4} and using Eq.~\eqref{intersection formula}, the surface elements $d^2\sigma_0$ and $d^2\sigma_1$ vanish, while $d^2\sigma_2$ reduces to $d\bar{t}d\bar{x}$.
Using the Fourier expansion in Eq.~\eqref{Fourier }, the derivative of the scalar field normal to the Wilson surface can be written as
\begin{equation}\begin{split}
		&\frac{\partial \gamma(x)}{\partial y}
		=\sum_{n=-\infty}^{\infty}\sum_{\vec{p}}
		ip_y e^{i(w_nt+\vec{p}\cdot\vec{x})}\gamma_n(\vec{p}),\\
		&\left.\frac{\partial \gamma(\bar{x})}{\partial y}\right|_{S}
		=\sum_{n=-\infty}^{\infty}\sum_{\vec{p}}
		ip_y e^{i(w_n\bar{t}+p_x\bar{x})}\gamma_n(\vec{p}).
		\label{partial gamma1}
\end{split}\end{equation}
Consequently, the surface integral takes the form
\begin{equation}\begin{split}
		\int_S d^2\sigma_\mu\partial_\mu\gamma(\bar{x}(\sigma))
		&=
		\sum_{n=-\infty}^{\infty}\sum_{\vec{p}}
		ip_y\gamma_n(\vec{p})
		\int_0^\tau d\bar{t}e^{iw_n\bar{t}}
		\int_0^{l_x}d\bar{x}e^{ip_x\bar{x}}\\
		&=
		\sum_{n=-\infty}^{\infty}\sum_{\vec{p}}
		ip_y\gamma_n(\vec{p})
		\tau l_x\delta_{w_n,0}\delta_{p_x,0}.
		\label{integral gamma}
\end{split}\end{equation}
At finite temperature, the Kronecker deltas $\delta_{w_n,0}\delta_{p_x,0}$ select the zero Matsubara mode and the momentum modes satisfying $p_x=0$. Therefore,
\begin{equation}
	e^{\frac{v^2\nu }{3\kappa}\int d^2\sigma_\mu\partial_\mu\gamma({\bar{x}})}
	=
	\prod_{\vec{p}}
	e^{i\frac{v^2\nu}{3\kappa}\tau l_xp_y\gamma_0\left(\vec{\mathtt{P}}\right)},
	\label{e2}
\end{equation}
where we define $\gamma_0\left(\vec{\mathtt{P}}\right)\equiv\gamma_0(0,\vec{p}_y)$.

Using Eqs.~\eqref{e1} and \eqref{e2}, the numerator of the fraction in Eq.~\eqref{W3} can be expressed in terms of the Fourier modes $\gamma_n(\vec{p})$. The resulting Gaussian integrations are evaluated using the standard relation
\begin{equation}
	\int_{-\infty}^{\infty} dx\, e^{-Ax^2+Bx}
	=\sqrt{\frac{\pi}{A}}\,e^{\frac{B^2}{4A}},
	\qquad A>0 .
	\label{Gaussian integral}
\end{equation}
Thus, the numerator becomes
	\begin{widetext}\begin{equation}\begin{split}
	\int_{-\infty}^\infty \left[D\gamma(x)\right] &e^{-\frac{v^2}{3\kappa}\int d^3x\left( \partial_\mu\gamma(x)\right) ^2} 
	e^{\frac{v^2\nu }{3\kappa}\int d^2\sigma_\mu\partial_\mu\gamma({\bar{x}}) }=
	\prod_{n=-\infty}^{\infty}\prod_{\vec{p}}\int_{-\infty}^\infty \left[ D\gamma_n(\vec{p})\right] \ e^{-\frac{v^2\mathtt{V}}{3\kappa}\left( w_n^2+\vec{p}^2\right)\left| \gamma_n(\vec{p})\right| ^2}e^{i\frac{v^2\nu}{3\kappa}\tau l_xp_y\gamma_0\left(\vec{\mathtt{P}}\right)}\\
	&=\left\lbrace \prod_{\overset{n=-\infty}{n\neq 0}}^{\infty}\prod_{\overset{\vec{p}}{p_x\neq 0}}\left( \dfrac{3\kappa\pi}{v^2\mathtt{V}}\right)^{\frac{1}{2}}\left[w_n^2+\vec{p}^2 \right] ^{-\frac{1}{2}}\right\rbrace
	\times\left\lbrace \prod_{\overset{\vec{p}}{p_x= 0}}\left( \dfrac{3\kappa\pi}{v^2\mathtt{V}}\right)^{\frac{1}{2}}\dfrac{1}{p_y}\right\rbrace \times e^{-\frac{v^2\nu^2}{12\kappa}\ \frac{\tau^2l^2_x}{\mathtt{V}}},
	\label{N}
	\end{split}
	\end{equation}\end{widetext}
Similarly, using Eq.~\eqref{e1} and the Gaussian integration formula given in Eq.~\eqref{Gaussian integral}, the denominator of the fraction in Eq.~\eqref{W3} is obtained as
	\begin{widetext}\begin{equation}\begin{split}
		\int_{-\infty}^\infty \left[D\gamma(x)\right] e^{-\frac{v^2}{3\kappa}\int d^3x\left( \partial_\mu\gamma(x)\right) ^2}
		&=\prod_{n=-\infty}^{\infty}\prod_{\vec{p}}\ \int_{-\infty}^\infty d\gamma_n(\vec{p})\ e^{-\frac{v^2\mathtt{V}}{3\kappa}\left( w_n^2+\vec{p}^2\right)\left| \gamma_n(\vec{p})\right| ^2 }\\
		&=\left\lbrace \prod_{\overset{n=-\infty}{n\neq 0}}^{\infty}\prod_{\overset{\vec{p}}{p_x\neq 0}} \left(\dfrac{3\kappa\pi}{v^2\mathtt{V}} \right) ^{\dfrac{1}{2}}\left[w_n^2+\vec{p}^2 \right] ^{-\dfrac{1}{2}}\right\rbrace
		\times\left\lbrace\prod_{\overset{\vec{p}}{p_x= 0}} \left(\dfrac{3\kappa\pi}{v^2\mathtt{V}} \right) ^{\dfrac{1}{2}}\dfrac{1}{p_y}\right\rbrace.
		\label{D}
	\end{split}
	\end{equation}\end{widetext}

	By substituting Eqs.~\eqref{N} and \eqref{D} into Eq.~\eqref{W3}, the average Wilson loop is obtained as
	\begin{equation}
			\left\langle W(C)\right\rangle=e^{-\frac{v^2 \nu^2}{12\kappa}\delta\left( \bar{y}\right)\times S}\times e^{-\frac{v^2\nu^2}{12\kappa}\ \frac{\tau^2l^2_x}{\mathtt{V}}}.
			\label{W4}
	\end{equation}
	In the limit $L\gg l_{x}$, with $L$ denoting the spatial length as defined earlier and the volume 
$\mathtt{V}=\beta L^2$,
	the second exponential term in Eq. \eqref{W4} reduces to unity, and the Wilson loop average $\left\langle W(C)\right\rangle$ is obtained,
	\begin{equation}
	\left\langle W(C)\right\rangle =e^{-\frac{v^2 \nu^2}{12\kappa}\delta\left( \bar{y}\right)\times S}.
	\label{Wdelta}
	\end{equation}
	Therefore, the area law fall-off for the Wilson loop which corresponds to a linear potential between a pair of static quark-antiquark, is obtained. In the next subsection, we obtain the string tension in terms of the intrinsic properties of the center vortices	and a parameter which describes the interaction between the vortices and discuss its behavior concerning the temperature. 

	\subsection{\label{subsec:results}Discussion on string tension and vortex properties  }

	In this subsection, we extract the string tension and discuss its characteristics in terms of the stiffness, the vortex tension, and the parameter describing the potential strength between the vortices. As mentioned in Section~\ref{subsec:vortex}, when a Wilson loop pierces the hypersurface $\Sigma$, it detects a singularity or discontinuity (see Fig.~\ref{fig1}.(a)), which results in a phase factor multiplying the Wilson loop (see Eq.~\eqref{phase difference}). On the other hand, since the linking number $L(C,l)$ is symmetric under the interchange of the loops $C$ and $l$, the same process can be interpreted from the viewpoint of the vortex loop. Namely, the Wilson surface $S$ is pierced by the vortex loop $l$, and the vortex loop detects the singularity located on the Wilson surface (see Fig.~\ref{fig1}.(c)).
	This singular structure appears as $\delta(\bar{y})$ in Eq.~\eqref{Wdelta}. The Dirac delta function originates from the definition of the vortex current $J^C_\mu(x)$ (see Eq.~\eqref{J}), which encodes the localization of the vortex field on the Wilson surface. In this sense, $\delta(\bar{y})$ reflects the intersection between the vortex configuration and the Wilson surface, rather than representing an independent contribution. Thus, this singular structure represents the discontinuity associated with the intersection of the vortex loop and the Wilson surface, in analogy with the discontinuity detected by the Wilson loop when it pierces the vortex configuration.
	
	We recall that, for simplicity, the Wilson surface is considered as a rectangular surface on the $t-x$ plane; therefore, the singularity appears on the $\bar{y}=0$ plane. Although center vortices are generally expected to possess a finite thickness or profile, the Dirac delta function appearing in the present derivation would still arise even if such a thickness or profile were taken into account. Rather, it originates from the intersection of the two-dimensional Wilson surface with the localized center-vortex configuration in the $t$--$x$ plane. The resulting singularity is therefore a direct consequence of this geometric intersection and reflects the localized interaction between the Wilson surface and the center vortex, rather than the internal structure of the vortex itself.
	
	To clarify the dimensional nature of this contribution, we note that although $S$ denotes the Wilson surface area with dimension $L^2$, the combination $\delta(\bar{y})S$ appearing in the exponent does not have the required dimension for an area term. In fact,
	\begin{align}
		\Big[\delta(\bar{y})\Big]\,\Big[S\Big]=\frac{1}{L}\times L^2=L .
	\end{align}
	Therefore, an appropriate dimensional analysis is required in order to identify the coefficient of this contribution as the physical string tension. We emphasize that the singular factor $\delta(\bar{y})$ alone does not define the string tension. To express this contribution in a dimensionally consistent form, a characteristic physical scale is required to construct dimensionless combinations from dimensionful quantities, which is a standard procedure in finite-temperature field theory. In the present formulation, we use the temperature scale $T$ for this purpose, since temperature naturally provides such a scale and explicitly enters the Fourier expansion through the Matsubara frequencies in Eq.~\eqref{Fourier }. Inserting the temperature scale through multiplication and division, the contribution can be recast in a dimensionally consistent form. Consequently, the combination $\delta(\bar{y})S/T$ acquires the dimension of an area, namely $L^2$, and the area-dependent contribution in the exponent of the Wilson-loop average takes the standard form required for the extraction of the physical string tension. 
	
	Therefore, the Wilson-loop average can be consistently rewritten as
	\begin{equation}
	\left\langle W(C)\right\rangle =e^{-\frac{v^2 \nu^2T}{12\kappa}\, \frac{\delta\left( \bar{y}\right)\times S}{T}}.
	\label{final W}
	\end{equation}
Consequently, the term $\dfrac{v^2 \nu^2}{12\kappa}T$ appears as the coefficient of the dimensionally consistent area contribution $\delta(\bar{y})S/T$, and it can therefore be interpreted as the string tension.

Replacing the vacuum expectation and the fact that for the fundamental representation $\nu=\dfrac{1}{2}$, the string tension is obtained as the following,
	\begin{equation}
	\sigma= \dfrac{ \left| \mu \right| \zeta}{48\kappa}T,
	\label{final string}
	\end{equation}
	where the sign of $\mu$ is applied. From Eq. \eqref{final string}, the relation between the tension $\left| \mu\right| $ and the stiffness $\dfrac{1}{\kappa}$ is,
	\begin{equation}
	\dfrac{1}{\kappa}=\dfrac{48\sigma}{\zeta T} \dfrac{1}{\left| \mu\right| }.
	\label{relation stiff and tension}
	\end{equation}
	This equation shows that the stiffness $\dfrac{1}{\kappa}$ decreases by increasing the tension $\left| \mu\right| $, in qualitative agreement with reference \cite{stiff1}.
	The reduction of the stiffness results in a more flexible vortex loop. 

	To plot the stiffness $\dfrac{1}{\kappa}$ versus the tension $\left| \mu\right| $, one must know the coefficient $\dfrac{48\sigma}{\zeta T}$ in Eq. \eqref{relation stiff and tension}. We have used figure $2$ of reference \cite{stiff1} and the coefficient is obtained to be approximately equal to 0.012. 
	It should be noted that the results of \cite{stiff1} were obtained in 4D space-time, whereas our calculations are performed in 3D. Strictly speaking, the infrared properties of 3D and 4D Yang–Mills theories are not identical in all respects; however, under certain conditions (e.g., dimensional reduction at high temperature), their long-distance behavior can be effectively related. Since our focus here is on qualitative features rather than precise numerical values of the parameters, we believe that a comparison with the results of \cite{stiff1} remains meaningful. We emphasize, however, that parameters such as $\zeta$ may have different dimensionalities in 3D and 4D, and therefore their numerical values should not be directly identified.
	
\begin{figure}[ht]
	\begin{center}
		\centering
		\includegraphics[height=7cm, width=8cm]{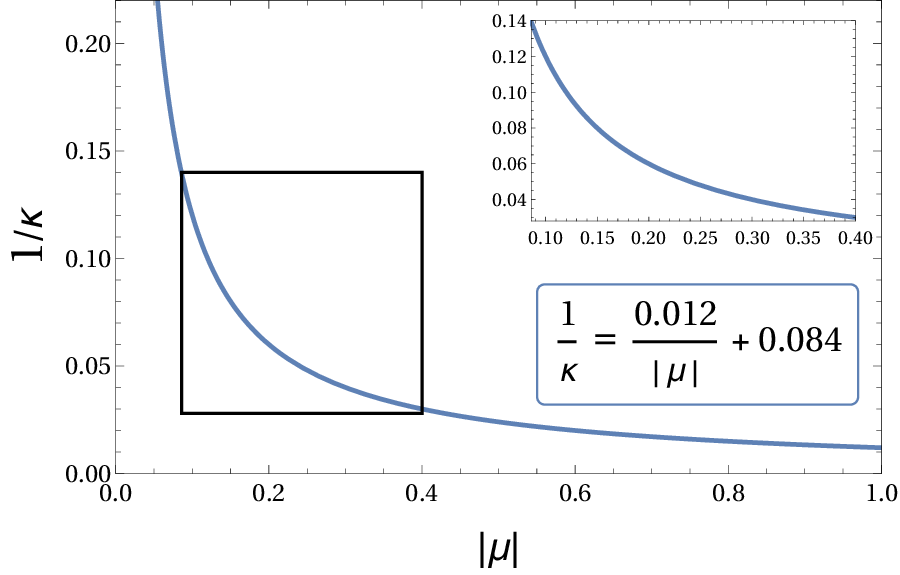}
		\caption{stiffness $\dfrac{1}{\kappa}$ versus the tension $\left| \mu\right| $. A magnification is done in the upper right of the plot to compare the data with the range of the data reported in Ref. \cite{stiff1}.}
		\label{fig5}
	\end{center}
	\end{figure}

	Using the coefficient $\dfrac{48\sigma}{\zeta T}$, Fig. \ref{fig5} plots $\dfrac{1}{\kappa}$ versus $\left| \mu\right| $ obtained from our calculations. The stiffness and the tension are scaled with the temperature and are dimensionless. A magnification is done in the upper right to compare the data with the range of the data reported in figure 2 of reference \cite{stiff1}.

	Since the parameters $\sigma$ and $\zeta$ are inherently dependent on temperature and using $\dfrac{48\sigma}{\zeta T}=0.012$ from the above discussion, the dimensionless string tension is rewritten from Eq. \eqref{relation stiff and tension},
	\begin{equation}
	\dfrac{\sigma(T)}{T^2}=\dfrac{0.012}{48} \dfrac{\zeta(T)}{T}.
	\label{sigma, zeta, T}
	\end{equation}

	\begin{figure}[ht]
	\begin{center}
		\centering
		\includegraphics[height=7cm, width=8cm]{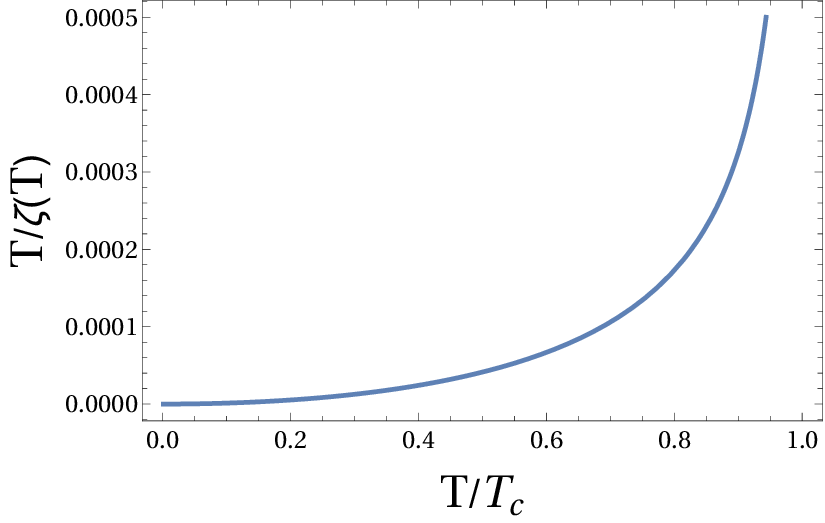}
		\caption{Dependence of the parameter which indicates the potential strength between the vortices, to the temperature.}
		\label{fig6}
	\end{center}
	\end{figure} 

	On the other hand, the temperature dependence of SU($2$) string tension was computed by lattice QCD \cite{zeta},
	\begin{equation}
	\dfrac{\sigma(T)}{T^2}=\dfrac{\sigma_0}{T^2_c}\left( \dfrac{T_c}{T}\right) ^2A\left( 1-\dfrac{T}{T_c}\right) ^{0.63}\left( 1+B\left( 1-\dfrac{T}{T_c}\right)^{\frac{1}{2}}\right),
	\label{stringT}
	\end{equation}
	where the constants $A=1.39$ and $B=\dfrac{1}{A}-1$ \cite{zeta}. The ratio $\dfrac{T_c}{\sqrt{\sigma_0}}\approx0.69$ with $\sigma_0=\left( 440\ \text{MeV}\right) ^2$ \cite{stiff1}, gives $T_c=303.6\ \text{MeV}$. From  Eqs. \eqref{sigma, zeta, T} and \eqref{stringT}, the dimensionless quantity $\dfrac{\zeta(T)}{T}$ is concluded,
	\begin{equation}
	\dfrac{\zeta(T)}{T}=\dfrac{48}{0.012}\dfrac{\sigma_0}{T^2}A\left( 1-\dfrac{T}{T_c}\right) ^{0.63}\left( 1+B\left( 1-\dfrac{T}{T_c}\right)^{\frac{1}{2}}\right),
	\label{zeta}
	\end{equation}
	as plotted in Fig. \ref{fig6}.
 	This plot shows that $T/\zeta$ increases with temperature, meaning the repulsive potential between vortices also increases as temperature rises.
	Increased repulsive forces can prevent condensation, which is expected in the deconfinement phase and the fact that by increasing the temperature one gets close to the deconfinement regime. Increasing the repulsive forces between the vortices at higher temperatures may prevent the vortices from having stable topological structures required for quark confinement.

\section{\label{sec:level5}Summary and Conclusions}

	Inspired by lattice results confirming the area law fall-off for the Wilson loop average, we calculate the Wilson loop average in the continuum limit for SU($2$) gauge group in three-dimensional Euclidean space-time. 

	We use the effective partition function obtained in Ref. \cite{oxman} for an ensemble of vortices containing complex scalar fields to calculate the Wilson loop average. We observe the area law fall-off for the Wilson loop and extract the string tension in terms of vortex parameters, including tension, stiffness, and the repulsive interaction parameter between vortices. Despite their discrete manner, we have employed an analytical and continuous approach—involving integration over compact scalar fields and the well-known Fourier expansion—to calculate the partition function.
	
	The stiffness-tension relation qualitatively agrees with the lattice results reported by \cite{stiff1}, where stiffness decreases with increasing tension. However, while their reduction is approximately linear, ours follows a $\frac{1}{|\mu|}$ dependence. By fitting our results to the lattice data from \cite{stiff1} and using the temperature-dependent string tension reported in \cite{zeta}, we obtain the temperature dependence of the repulsive interaction parameter between vortices. Our result demonstrates that the repulsive force between the vortices increases with the temperature. This behavior is consistent with expectations from vortex models, where increased repulsion at higher temperatures would lead to reduced vortex effects. It is also in agreement with lattice results where the effect of topological defects must be decreased at smaller distances.

\end{document}